\documentclass[12pt]{article}
\usepackage{graphicx}
\usepackage{natbib} 
\usepackage{url} 
\usepackage{amsmath, amsfonts, amssymb, amsthm, natbib,bm,enumerate,multirow}
\usepackage{psfrag,epsf,array, adjustbox}
\usepackage{latexsym}
\usepackage{calc}
\usepackage[ragged, symbol]{footmisc}
\usepackage{grffile}
\usepackage{booktabs}
\usepackage{caption}
\usepackage{comment}
\usepackage{color,url}

\newcommand{\half}{\frac{1}{2}}
\newcommand{\sigsq}{\sigma^2}

\newcommand{\ident}{\bm{I}}

\newcommand{\bX}{\bm{X}}
\newcommand{\bx}{\bm{x}}
\newcommand{\bZ}{\bm{Z}}
\newcommand{\bz}{\bm{z}}
\newcommand{\bu}{\bm{u}}

\newcommand{\bbeta}{\bm{\beta}}
\newcommand{\bpsi}{\bm{\psi}}
\newcommand{\bxi}{\bm{\xi}}

\newcommand{\Ytilde}{\widetilde{Y}}
\newcommand{\Xtilde}{\widetilde{X}}
\newcommand{\inv}{^{-1}}

\newcommand{\bV}{\bm{V}}

\newcommand{\btheta}{\bm{\theta}}
\newcommand{\superth}{^{th}}
\newcommand{\bP}{\bm{P}}

\newcommand{\blind}{0}

\begin{document}

\def\spacingset#1{\renewcommand{\baselinestretch}%
{#1}\small\normalsize} \spacingset{1}

\if0\blind
{
	\title{\bf Model Testing for Generalized Scalar-on-Function Linear Models}
	\author{Stephanie T. Chen\thanks{
	The authors gratefully thank Dr. Daowen Zhang and Dr. Matthew McLean for providing code to conduct the \textit{aScore} and likelihood ratio test for functional generalized additive models, respectively. ST Chen and L Xiao's research were supported by grant numbers OPP1148351 and OPP1114097 from the Bill and Melinda Gates Foundation. AM Staicu's research was supported by NSF grant number DMS 1454942 and NIH grants 5P01 CA142538-09 and 2R01MH086633. This work represents the opinions of the researchers and not necessarily that of the granting organizations.}\hspace{.2cm}\\
		Department of Statistics, North Carolina State University\\
		and \\
		Luo Xiao \\
		Department of Statistics, North Carolina State University\\
		and \\
		Ana-Maria Staicu\\
		Department of Statistics, North Carolina State University}
	\maketitle
} \fi

\if1\blind
{
	\bigskip
	\bigskip
	\bigskip
	\begin{center}
		{\LARGE\bf Model Testing for Generalized Scalar-on-Function Linear Models}
	\end{center}
	\medskip
} \fi

\bigskip
\begin{abstract}
	Scalar-on-function linear models are commonly used to regress functional predictors on a scalar response. However, functional models are more difficult to estimate and interpret than traditional linear models, and may be unnecessarily complex for a data application. Hypothesis testing can be used to guide model selection by determining if a functional predictor is necessary. Using a mixed effects representation with penalized splines and variance component tests, we propose a framework for testing functional linear models with responses from exponential family distributions.
	The proposed method can accommodate dense and sparse functional data, and be used to test functional predictors for no effect and form of the effect. We show via simulation study that the proposed method achieves the nominal level and has high power, and we demonstrate its utility with two data applications.
\end{abstract}

\noindent%
\vfill

\newpage
\spacingset{2} 
\section{Introduction}\label{sec:Introduction}
The rapid rise in computing and storage capabilities has led to increasing availability of continuous and infinite-dimensional functional data, in fields such as medicine, economics, and signal processing (see \cite{RamsaySilverman2002} for an overview). Functional linear models, which extend standard linear models to include functional predictors and/or responses, are one of the most common methods for analyzing functional data (see \cite{Morris2015} for a recent review). We focus on models with functional predictors and a scalar response, commonly called scalar-on-function linear models (Chapter 15 in \cite{RamsaySilverman2005}, Chapter 1 in \cite{FerratyVieu2006}). By incorporating the full functional predictor (rather than a simple summary statistic), functional models can significantly improve model performance, but are also much more difficult to estimate and interpret. Hypothesis testing can be used to determine the necessity of a functional data model. In this paper, we propose a new test for scalar-on-function linear models with generalized responses.

This work is motivated by a dataset of diffusion tensor imaging (DTI) of intracranial white matter for patients with multiple sclerosis (MS), a neurodegenerative disease characterized by damage to the myelin sheath that causes degradation of physical and mental ability (see \cite{Reichetal2010} for study details). DTI measures the diffusion of water in the brain and can be used to map demyelination of white matter. These DTI scans are summarized as profiles that measure a magnetic resonance imaging (MRI) index, such as mean diffusivity or fractional anisotropy, as a function of location along white matter tracts. Many studies use functional models to map the relationship between tract profiles and MS status or disability progression \citep{Gertheissetal2013, Goldsmithetal2011, Goldsmithetal2012, Ivanescuetal2015, Zhuetal2010}. We focus on the study in \cite{Goldsmithetal2011} who attempt to identify patients with MS using logistic regression models that include (a) no tract profile information, (b) tract profile average only, and (c) full tract profile as a functional predictor. Hypothesis testing can provide a scientifically rigorous approach to determine which tract profiles are related to MS. Additionally, testing can determine if modeling the full tract profile is significantly better than simply modeling its average, that is, comparing standard versus functional logistic regression models. To the authors' knowledge, there are no existing methods for binary (generalized) responses.

A number of testing approaches exist for functional models with Gaussian responses. \cite{Cardotetal2003a} develop a permutation-based test and \cite{Kongetal2016} extend Wald, score, likelihood ratio, and F-tests to test if a functional predictor relates to the response (nullity). \cite{Swihartetal2014} and \cite{McLeanetal2015} use penalized splines to estimate the functional effect with a mixed model representation, and then frame hypothesis tests in terms of random and/or fixed effects in the mixed model. Methods for testing mixed models exist for Gaussian responses, such as the finite-sample likelihood ratio test (LRT) by \cite{CrainiceanuRuppert2004}, but are limited for generalized responses \citep{Lin1997, MolenberghsVerbeke2007, Chenetal2019}. Thus, existing approaches for testing scalar-on-function linear models cannot be extended to generalized responses, and there is need for new methodology.

We propose a unified testing framework for association between a functional predictor and generalized response in a scalar-on-function linear model. We consider testing for (a) no association between predictor and response and (b) for a specific polynomial form of the association, e.g. a constant form. Similar to \cite{Swihartetal2014} and \cite{McLeanetal2015}, we exploit the mixed model representation of penalized splines to present hypothesis testing in terms of a generalized linear mixed model (GLMM). In contrast to existing approaches, our method adaptively chooses the spline penalty such that testing is always conducted as a restricted likelihood ratio test (RLRT) of a single variance component. We will show that this proposed framework has better performance than existing method for normal and generalized responses.

The remainder of this paper is organized as follows. Section 2 introduces the statistical model and hypothesis tests. Sections 3 and 4 present the proposed method and its implementation, respectively. Section 5 presents a simulation study and Section 6 describes two data applications, including the motivating example. Finally, Section 7 summarizes the paper.
\section{Statistical Framework}\label{sec:Methods}
Consider observed data $\{Y_i,(X_{i}(t_{ij}), t_{ij}): i=1,\dots,n; j = 1,\dots,m_i; t_{ij}\in \mathcal{T}\}$, where $Y_i$ is a scalar response for the $i\superth$ subject from an exponential family distribution and $X_{i}(t)$ is the observed functional predictor measured at $t_{ij}$, the $i\superth$ subject's $j\superth$ point. For clarity, we assume that $X_i(t)$ has zero mean and is observed without noise; we will discuss relaxing these assumptions in Section 4.
Without loss of generality, let $\mathcal{T} \in [a,b]$ be a compact domain. We will consider noisy functional predictors in a later section. Our goal is to characterize the relationship between functional predictor, $X_{i}(t)$, and scalar response, $Y_i$. To do so, we will pose a generalized functional linear model (GFLM) and use hypothesis testing to determine the form of the relationship.
We posit the GFLM for $Y_i$ as
\begin{equation}
	\begin{split}
		E(Y_i|X_i) &=  g^{-1}(\eta_i)
		\\  \eta_i &=\alpha+\int_{\mathcal{T}} X_i(t) \beta(t)dt,
	\end{split}
\end{equation}
where $g(x)$ is a known link function, $\eta_i$ is the linear predictor for the $i\superth$ subject, $\alpha$ is a fixed intercept, and $\beta(t)$ is a differentiable smooth coefficient function that weights $X_i(t)$ over $t$. We are interested in the form of $\beta(t)$. For example, if $\beta(t)=0$, then the predictor has no effect on $Y_i$. If $\beta(t)= \beta$ for some scalar $\beta$, then $\int_{\mathcal{T}} X_i(t)\beta(t) dt = \beta\bar{X}_i$ and the model reduces to a generalized linear model. This information about $\beta(t)$ can guide development of an appropriate data-driven model to maximize interpretability and computational efficiency. Note that $\beta(t)$ is only identifiable up to the span of $X_i(t)$ such that the part of $\beta(t)$ that is orthogonal to $X_i(t)$ is not estimable \citep{Cardotetal2003b}; see \cite{ScheiplGreven2016} for detailed justification.

We consider three hypothesis tests for specific forms of the smooth coefficient function:
\begin{enumerate}[(a)]
	\item \textbf{No effect of the predictor (nullity)}
	\\$H_0 \text{: } \beta(t) = 0 \: \forall \: t$ vs $H_A \text{: } \beta(t) \ne 0 $
	\item \textbf{Necessity of the functional covariate (functionality)}
	\\$H_0 \text{: } \beta(t) = \beta_0 \: \forall \: t$ for some scalar $\beta_0$ vs $H_A \text{: } \beta'(t) \ne 0 $ 
	\item \textbf{Linearity of the smooth coefficient (linearity)}
	\\ $H_0 \text{: } \beta(t) = \beta_0+\beta_1t$ for some scalars $\beta_0$, $\beta_1$ vs $H_A \text{: } \beta''(t) \ne 0$, 
\end{enumerate}
where $\beta'(t)$ and $\beta''(t)$ denote the first and second derivatives of $\beta(t)$. These specific forms have scientific and computational implications. For example, in the DTI application considered in the literature, the test for nullity checks if a tract profile has any relationship with MS disease status. Rejecting $H_0$ indicates that the tract is useful for detecting MS. The test for functionality investigates if modeling the full tract profile is actually necessary. Rejecting $H_0$ indicates that modeling just the tract mean is not sufficient, and that a functional linear model is necessary. Finally, the test for linearity checks a commonly assumed coefficient form, and can also be used to compare a functional linear model with a functional generalized additive model \citep{McLeanetal2015}. Rejecting $H_0$ indicates that a bivariate additive model improves on the standard functional linear model. These hypotheses have been extensively considered for functional models with Gaussian responses, but not for generalized responses. For example, \cite{Kongetal2016} consider classical tests for testing nullity, \cite{Swihartetal2014} test for nullity and functionality, and \cite{McLeanetal2015} test for nullity and linearity. While these methods are effective, each hypothesis requires a different modeling framework. In the following section, we propose a unified testing framework for all three hypotheses for functional linear models with generalized responses. 

\section{Methodology}
\subsection{Outline}
We propose a unified framework for testing the smooth coefficient in a scalar-on-function linear model with generalized responses. Our method consists of three main steps: (a) use penalized splines and functional principal components to approximate equation (1) with a GLMM, (b) re-frame the hypothesis tests from Section 2 in terms of a single zero-value variance component, (c) use the approximate restricted likelihood ratio test (RLRT) from \cite{Chenetal2019} to test the variance component.

\subsection{Preliminary Generalized Functional Linear Model}
We begin by approximating the GFLM described in equation (1) as a GLMM using functional principal components for the predictor, $X_i(t)$, and a penalized spline representation for the coefficient, $\beta(t)$. First, take the spectral decomposition of the covariance of $X_i(t)$ as $\sum_{k=1}^\infty \lambda_k \psi_k(t)\psi_k(t')$, where $\lambda_k\ge 0$ is the $k\superth$ eigenvalue corresponding to eigenfunction $\psi_k(t)$. We assume that $X_i(t)$ can be accurately approximated by a finite basis expansion, and use a truncated Karhunen-Lo\'{e}ve approximation 
\begin{equation}
	X_i(t) \approx \sum_{k=1}^{K_x} \xi_{ik}\psi_k(t) = \bxi_i^T\bpsi(t),
\end{equation}
for a $K_x$ finite truncation. This is a common assumption in functional data analysis to reduce the dimensionality of $X_i(t)$ \citep{RamsaySilverman2005, Yaoetal2005}, and accommodates scenarios where the predictor may be sparsely observed or noisy. Here, $\xi_{ik}=\int X_i(t)\psi_k(t)dt$ is the $i\superth$ subject's score for the $k\superth$ eigenfunction, and let $\bxi_i=[\xi_{i1},\dots, \xi_{iK_x}]^T$ and $\bpsi(t)= [\psi_1(t), \dots, \psi_{K_x}(t) ]^T$. These components can be estimated by computing the sample covariance of $X_i(t)$ and estimating its spectral decomposition \citep{Yaoetal2005, Xiaoetal2016}. We will discuss selecting the truncation parameter, $K_x$, in Section 4.

Next, we approximate the coefficient function, $\beta(t)$, with B-spline basis, $B_k(t)$, where $\bm{B}(t)=[B_1(t),\dots,B_{K_u}(t)]^T$ is a vector of splines evaluated at $t$, corresponding to random basis coefficients $\bm{g}=[g_1,\dots,g_{K_u}]^T$, such that $\beta(t) \approx \sum_{k=1}^{K_u} g_k B_k(t)$. Here, $K_u$ must be large enough to capture the complexity of $\beta(t)$. To impose smoothness, we treat $\bm{g}$ as random effects with a $d\superth$-order differencing penalty matrix, $\bm{P}_d$, which has rank $(K_u - d)$ \citep{EilersMarx1996}. Following \cite{Goldsmithetal2012}, we can transform $\bm{g}$ into a set of penalized and unpenalized terms using an eigendecomposition of $\bm{P}_d$. Specifically, let $\bm{P}_d=\bm{Q\Lambda Q^T}$ such that $\bm{Q}=[\bm{Q}_1, \bm{Q}_2]$ is a matrix of orthonormal eigenvectors and $\Lambda$ is the diagonal matrix of corresponding eigenvalues where $\bm{\Lambda}_1$ is the matrix of ($K_u - d$) non-zero eigenvalues. Define $\bm{Q}_1$ as the set of eigenvectors corresponding to $\Lambda_1$, and $\bm{Q}_2$ as the remaining $d$ eigenvectors. Then, we can define $\bu^* = \bm{Q}_1^T\bm{g}$ as the $(K_u - d)$ length vector of penalized terms, where $\bu^* \sim N(0,\sigsq_u\bm{\Lambda}_1^{-1})$, and $\bbeta^* = \bm{Q}_2^T\bm{g}$ as the $d$-length vector of unpenalized terms. Using this formulation, $\beta(t)$ can be approximated as
\begin{equation}
	\beta(t) \approx \sum_{k=1}^{K_u}g_k B_k(t) = \bm{B}(t)^T\bm{g}=\bm{B}(t)^T[\bm{Q}_1\bu^*+\bm{Q}_2\bbeta^*].
\end{equation}
In this context, the smooth function is defined by the choice of $\bm{P}_d$. For example, if $\bm{P}_2$ is a second-order differencing penalty, then $\beta(t)$ is a linear function if and only if $\bm{u}^* = \bm{0}$ or equivalently $\sigsq_u = 0$ \citep{EilersMarx1996}. In general, if $\bP_d$ is a $d\superth$ order penalty, then $\beta(t)$ is a $(d-1)$ degree polynomial if and only if $\sigsq_u = 0$. 

By substituting the approximations for $X_i(t)$ and $\beta(t)$ into equation (1), we can approximate the linear predictor, $\eta_i$, with a GLMM. Let $\bm{J}=\int_{\mathcal{T}} \bpsi(t) \bm{B}(t)^T dt$ be the $K_x \times K_u$ matrix of integrated products of the eigenfunction and B-spline pairs. Then
\begin{equation}
	\begin{split}
		\eta_i &\approx \alpha + \bm{\xi}_i^T\bm{JQ_2}\bbeta^* + \bm{\xi}_i^T\bm{JQ_1}\bu^*
		\\ &= \bm{x}_i^T\bbeta + \bm{z}_i^T\bu,
	\end{split}
\end{equation}
where $\bm{x}_i^T= (1, \bm{\xi}_i^T\bm{JQ}_2)$ is the row vector corresponding to fixed effects $\bm{\beta}=(\alpha, \bbeta^{*^T})^T$ and $\bm{z}_i^T=\bm{\xi}_i^T\bm{JQ}_1\bm{\Lambda}_1^{-\half}$ is the row vector corresponding to random effects $\bu=\bm{\Lambda}_1^{\half}\bu^*\sim N(0,\sigsq_u\ident)$. By collecting all terms over $i$, we can write equation (4) in matrix form as $\bm{\eta} \approx \bX\bbeta + \bZ\bu$, where $\bm{\eta}$, $\bX$ and $\bZ$ are the respective row-stackings of $\eta_i$, $\bm{x}_i^T$, and $\bm{z}_i^T$. Note that when $d=0$, then $\bm{x}_i^T = 1$ and $\bm{\beta} = \alpha$. 
With this formulation, testing the form of $\beta(t)$ can be re-framed as 
testing if $\sigsq_u = 0$ with appropriate choice of $\bm{P}_d$.

\subsection{Hypothesis Testing}
In the second step, we present the hypothesis framework as a test of $H_0: \sigsq_u$ versus $H_A: \sigsq_u > 0$ in equation (4) and choice of the differencing penalty matrix, $\bm{P}_d$:
\begin{enumerate}[(a)]
	\item \textbf{Nullity:} $\bm{P}_0$ is zero-order ($d=0$)
	\item \textbf{Functionality:} $\bm{P}_1$ is first-order ($d=1$)
	\item \textbf{Linearity:} $\bm{P}_2$ is second-order ($d=2$).
\end{enumerate}
Our proposed framework selects the penalty matrix, $\bm{P}_d$, to ensure that the hypothesis test is always in terms of a single variance component. This differs from the frameworks used by \cite{Swihartetal2014} ($\bm{P}_d = \bm{P}_1$) and \cite{McLeanetal2015} $\bm{P}_d = \bm{P}_2$ that fix $d$ and test different effects for each hypothesis. For example, using the \cite{Swihartetal2014} framework, testing for functionality involves a single random effect while testing for nullity is more challenging, and involves a fixed and random effect. The choice of $\bm{P}_d$ can also impact identifiability of the smooth coefficient. In practice, \cite{ScheiplGreven2016} note that first-order difference penalties avoid most identifiability issues, but the chance of non-identifiability increases for higher-order differences. Thus, testing for linearity or higher level polynomials may be more challenging and have worse performance than testing for nullity or functionality.

\subsection{Approximate RLRT for a Generalized Linear Mixed Model}
To test the hypotheses in Section 3.3, we will use the RLRT
\begin{equation}
	\textit{RLRT} = -2 \Big\{\sup_{\btheta \in H_0} \widetilde{\text{REL}}(\btheta) - \sup_{\btheta \in H_A} \widetilde{\text{REL}}(\btheta) \Big\},
\end{equation}
where $\widetilde{\text{REL}}(\btheta)$ denotes the restricted log-likelihood of equation (4) and $\btheta = (\bbeta^T, \sigsq_u)^T$. When available, the RLRT is shown to outperform the LRT for normal responses \citep{Scheipletal2008}. Numerical results corroborate this observation for generalized responses \citep{Chenetal2019}. However, the RLRT is only appropriate when the fixed effects are identical under the null and alternative hypotheses. Thus, we expect the proposed adaptive framework to outperform existing methods that require simultaneous testing of fixed and random effects.

For our hypothesis tests, the null distribution of the LRT and RLRT is non-standard because the null value lies on the boundary of the parameter space. \cite{SelfLiang1987} derive the asymptotic null distribution as a mixture of chi-square distributions, and \cite{MolenberghsVerbeke2007} propose an asymptotic LRT for GLMMs using this result. However, this asymptotic distribution leads to conservative tests for normal and generalized responses \citep{CrainiceanuRuppert2004, PinheiroBates2000, Chenetal2019}. While \cite{CrainiceanuRuppert2004} derive the finite sample distribution for normal responses and show that it outperforms the asymptotic distribution, no such results exist for generalized responses.

Instead, we use the approximate RLRT developed by \cite{Chenetal2019} for testing variance components in GLMMs. This method approximates the RLRT for a GLMM with the RLRT for a working linear mixed model (LMM) by extending the penalized quasi-likelihood (PQL) approach for parameter estimation \citep{Schall1991, BreslowClayton1993, WolfingerOconnel1993}. Briefly, PQL estimation consists of two iterative steps: (a) calculation of a ``normalized" working response, $\Ytilde_i$, and (b) estimation of a working LMM for $\Ytilde_i$. At convergence, define $\Ytilde_i = W_i^{*\half}[\eta_i^*+ g'(\mu_i^*)(Y_i-\mu_i^*)]$, where $\eta_i^*$ is the linear predictor for the GLMM, $g'(\mu_i^*)$ is the derivative of the link function evaluated at the conditional mean, $\mu_i^*$, and $W_i^* = \big[h_i^2 V_i^* \big]\inv$ weights observations with $h_i = g'(\eta_i^*)$ and $V_i^* = Var(Y_i|\bu)$. Then, the working LMM corresponding to equation (4) is $\Ytilde_i \approx \widetilde{\bx}_i^T\bbeta + \widetilde{\bz}_i^T\bu + e_i$, where $\widetilde{\bx}_i$ and $\widetilde{\bz}_i$ are $\bx_i$ and $\bz_i$ left-multiplied by $W_i^*$, $\bbeta$ and $\bu \sim N(\bm{0},\sigsq_u \ident)$ are as defined in equation (4), and $e_i \sim N(0,\sigsq_e)$ for all $i$. This allows for use of the finite-sample null distribution derived by \cite{CrainiceanuRuppert2004}. \cite{Chenetal2019} show that this finite-sample approximate RLRT outperforms the asymptotic LRT applied directly to the GLMM. Using this approach, the $\widetilde{\text{REL}}$ in equation (5) can be calculated for the vector of responses, $\bm{\Ytilde} = (\Ytilde_1, \dots, \Ytilde_n)^T$, as $\widetilde{\text{REL}}(\btheta) = -\half \Big[\log|\widetilde{\bV}| + \log| \widetilde{\bX}^T\widetilde{\bV}\widetilde{\bX}| + (n-p)\log(\bm{\Ytilde}^T\widetilde{\bP}^T\widetilde{\bV}\inv\widetilde{\bP}\bm{\Ytilde}) \Big]$, where $\bm{\Xtilde}$ is the row-stacking of $\widetilde{\bx}_i^T$,  $\widetilde{\bV} = Var(\bm{\Ytilde})$ is the marginal variance, $\widetilde{\bP} = \ident - \bm{\Xtilde}^T(\bm{\Xtilde}^T\widetilde{\bV}\inv\bm{\Xtilde})\inv \bm{\Xtilde}\widetilde{\bV}\inv$ is a projection matrix, and $p$ is the dimension of $\bbeta$. The resulting test statistic can then be compared to its finite-sample distribution.

\section{Implementation}
To approximate $X_i(t)$ in equation (3), we use the fast covariance estimation method \citep{Xiaoetal2016} implemented in the \texttt{fpca.face} function in \texttt{R} package \texttt{refund} \citep{refund} with default settings. This approach can accommodate functional predictors with non-zero mean and noisy observations by de-meaning the predictor using smoothing splines, and smoothing the resulting sample covariance \citep{Xiaoetal2016}.
To estimate the truncation parameter, $K_x$, we use the Aikaike Information Criterion (AIC) by \cite{Lietal2013}. For functional data, AIC is given as $\arg \min_k N\log(\hat{\sigma}^2_{[k]})+N+2nk$, where $N$ is the total number of observations, $n$ is the number of subjects, and $\hat{\sigma}^2_{[k]}$ is the marginal error variance using $k$ eigenfunctions. The $\hat{\sigma}^2_{[k]}$ can be calculated by integrating the difference between the full error variance and diagonal of the marginal covariance matrix generated using $k$ eigenfunctions (Li et al., 2013). The test for linearity requires a minimum of three eigenfunctions for estimation, so we let $K_x$ be the larger of three and the parameter selected by AIC. In general, $K_x\ge d+1$ is needed to test higher-order polynomials.

To model $\beta(t)$, we follow \cite{Swihartetal2014} and use $K_u=30$ cubic B-splines with equally-spaced knots. To estimate the GLMM in equation (4), we use penalized quasi-likelihood with the \texttt{glmmPQL.mod} function and conduct the approximate RLRT with function \texttt{test.aRLRT}; both are available in the \texttt{glmmVCtest} package \citep{glmmVCtest}.

\section{Simulation Study}\label{sec:Simulation Study}
We conduct a simulation study to evaluate performance of the proposed method, referred to as the \textit{aRLRT} method, and five extensions of existing approaches, described in Section 5.1. Generate the generalized responses, $Y_i$, and the functional predictor, $X_{i}(t)$, as
\begin{equation}
	\begin{split}
		Y_i &= g^{-1}(\eta_i)
		\\ \eta_i &= \alpha + \int X_i(t)\beta(t)dt
		\\ X_{i}(t_{ij}) &= \mu(t_{ij})+\sum_{k=1}^5 \xi_{ik}\psi_k(t_{ij}) + \epsilon_{ij},
	\end{split}
\end{equation}
where $g(x)$ is the canonical link function, $\alpha=0$, $\xi_{ik} \sim N(0,\lambda_k)$, $\epsilon_{ij}\sim N(0,\sigsq_X)$, and $\mu(t)$, $\lambda_k$, $\psi_k(t)$, and $\sigsq_X$ are estimated from the baseline corpus callosum (CCA) profiles for multiple sclerosis patients in the DTI dataset, available in the \texttt{refund} package \citep{refund}. That is, use the \texttt{fpca.face} function to estimate $\mu(t)$, the first five eigenvalues, $\lambda_k$, and eigenfunctions, $\psi_k(t)$, and the measurement error, $\sigsq_X$. Let $t_{ij}$ be observed on a grid of 80 equally-spaced points from $[0,1]$. If $m_i = 80$, the subject is observed at all points and if $m_i < 80$, observed points are uniformly sampled for each subject from the 80 possible points. Consider a factorial combination of the factors:
\begin{enumerate}
	\item \textbf{Distribution of $Y_i$:} (a) Bernoulli, (b) Normal, (c) Binomial, (d) Poisson
	\item \textbf{Form of $\beta(t)$:} (a) Scalar: $ \delta_0$, (b) Linear: $1+\delta_1 t$, (c) Trigonometric: $ 1+t+\delta_2\cos(2\pi t) $
	\item \textbf{Number of subjects} ($i = 1, \dots, n$): (a) $n=100$, (b) $n=500$
	\item \textbf{Observations per subject} ($j = 1, \dots, m_i$): (a) $m_i=80$ (dense), (b) $m_i=40$, (c) $m_i=20$, (d) $m_i=10$
\end{enumerate}
In the forms of $\beta(t)$, $\delta_0$, $\delta_1$, and $\delta_2$ are scalar coefficients controlling deviation from the null hypothesis. For each setting, we consider 5000 simulated datasets at the $\alpha=0.05$ level for type I error and power. For conciseness, only results for dense Bernoulli data are shown in the main text. Results for non-dense Bernoulli data, and Poisson, Normal, and Binomial distributions are included in the Supplemental Materials. 

\subsection{Alternative Methods}
\subsubsection{Approximate Score Test (\textit{aScore})}
\cite{Lin1997} and \cite{ZhangLin2003} develop a score test for variance components in GLMMs.
We consider a variant of the proposed method by testing the hypotheses in Section 3.3 using the approximate bias-corrected score test from \cite{ZhangLin2003}, and refer to this method as \textit{aScore}.

\subsubsection{Approximate Penalized Functional Regression (\textit{aPFR})}
The penalized functional regression (PFR) framework by \cite{Swihartetal2014} uses a mixed model representation induced by a modified first-order penalty.
Testing for functionality involves a single random effect and can be extended to generalized responses using the approximate RLRT by \cite{Chenetal2019}. The test for nullity requires simultaneous testing of a fixed and random effect. To extend this method to generalized responses, we modify the approximate test by \cite{Chenetal2019} to use likelihood instead of restricted-likelihood, thus becoming an approximate LRT. The test for linearity cannot be conducted. We refer to this full framework as the \textit{aPFR} method. Note that LRTs generally have lower power than RLRTs \citep{Scheipletal2008, Chenetal2019}, so we expect this method to have worse performance for testing nullity compared to the proposed \textit{aRLRT} method.

\subsubsection{Functional Principal Components Regression (\textit{FPCR})}
The functional principal components regression (FPCR) framework from \cite{Swihartetal2014} represents a scalar-on-function model with a linear model to frame hypothesis tests in terms of fixed effects. Note that because this framework centers the functional predictor by subject, it can be used to test nullity and functionality, but not linearity. This approach uses a standard LRT for testing so can be applied to generalized responses without modification, and we refer to it as \textit{FPCR}.

\subsubsection{Approximate Functional Generalized Additive Model (\textit{aFGAM})}
The method by \cite{McLeanetal2015} uses a functional generalized additive model (FGAM) framework to present testing in terms of a mixed model induced by a second-order penalty. Their test for linearity involves a single random effect, while the tests for functionality or nullity involve simultaneous testing of a random effect with one or two fixed effects, respectively. To extend this method to generalized responses, we use the approximate RLRT \citep{Chenetal2019} and approximate LRT discussed for the \textit{aPFR} method. We refer to this framework as \textit{aFGAM}. Again, we expect the \textit{aFGAM} method to have inferior performance for testing nullity and functionality compared to the proposed \textit{aRLRT} method.

\subsubsection{Approximate F-test (\textit{aFtest})}
\cite{Kongetal2016} use a FPCR framework to present testing for nullity in terms of fixed effects in a linear model, and conduct hypothesis testing using a standard F-test. Unlike the \textit{FPCR} method previously discussed, the functional predictors are centered over all subjects and the framework cannot be used to test for functionality or linearity. To extend this approach to generalized responses, we can apply the F-test to the ``normalized" working responses from PQL estimation (as discussed in Section 3.4). We refer to this method as \textit{aFtest}.

\subsection{Simulation Results}
For brevity, only results for the \textit{aRLRT} and \textit{aScore} methods for Bernoulli responses with densely observed functional predictors are included in the main text; all others are in the Supplementary Materials.

Table 1 reports the empirical type I error rates for testing binary responses with dense functional predictors. For the settings considered, both the \textit{aRLRT} and \textit{aScore} methods have rates close to the nominal $\alpha=0.05$ level for all three hypotheses. Both methods have type I error rates generally close to the nominal level for moderate levels of noise and sparsity, but can become conservative as $X_i(t)$ is less accurately estimated (Supplemental Table S1). In comparison, the \textit{aPFR} and \textit{aFGAM} methods maintain nominal levels when testing involves only random effects (Table S2; functionality for \textit{aPFR}, linearity for \textit{aFGAM}), but are conservative for $n=100$ when the hypothesis involves simultaneous testing of fixed and random effects. This shows the merits of an adaptive framework that allows for testing using RLRTs rather than LRTs. The \textit{FPCR} method is inflated for $n=100$, and the \textit{aFtest} method is inflated for all settings; both methods improve with sample size (Table S2). Additionally, the type I error rates for all methods can be inflated for testing functionality and nullity when the magnitude of the smooth coefficient is very large (results not shown). In these scenarios, the probability of a Bernoulli event is converging to 0\% or 100\%, making estimation and testing of a logistic regression model unsuitable for the data. This issue occurs only for Bernoulli responses. For testing Normal, Binomial, and Poisson responses, type I error rates are generally close to nominal levels for all methods except \textit{aFtest}. The \textit{aFtest} is inflated for Binomial responses with $n=100$ and Poisson responses for all settings.

\begin{table}
	\begin{center}
	\caption{Empirical type I error rates for binary responses using the \textit{aRLRT} and \textit{aScore} methods at the nominal $\alpha=0.05$ level based on 5000 datasets, by number of subjects, $n$, and form of the smooth coefficient, $\beta(t)$. Only rejection probabilities corresponding to type I errors are displayed. The maximum standard error was 0.0035.}
	\scalebox{0.9}{\begin{tabular}{ c | c |c | c c c | c c c }
			\multicolumn{3}{c|}{} & \multicolumn{3}{c|}{\textit{aRLRT}} & \multicolumn{3}{c}{\textit{aScore}} \\ \hline
			$n$ & $\beta(t)$ & $\delta_\ell$ & Linearity & Functionality & Nullity & Linearity & Functionality & Nullity \\ \hline
			\multirow{6}{*}{100} &\multirow{3}{*}{$\delta_0$} &
			0 & 0.048 &	0.046 &	0.043 &	0.054 &	0.052 &	0.051 \\
			& & 5 & 0.046 & 0.039 & &	0.051 &	0.045 & \\
			& & 10 & 0.046 &	0.042 & &	0.050 &	0.046 &  \\
			& \multirow{3}{*}{$1+\delta_1 t$} & 0 & 0.048 &	0.041 & &	0.052 &	0.047 & \\
			& & 5 & 0.048 & & & 0.054 & & \\
			& & 10 & 0.049 & & & 0.054 & &	 \\	
			\hline
			\multirow{6}{*}{500} & \multirow{3}{*}{$\delta_0$}  & 
			0 & 0.053 &	0.044 & &	0.050 &	0.045 & \\
			& & 5 & 0.054 &	0.044 &	& 0.051 &	0.043 & \\
			& & 10 & 0.053 &	0.049 & & 0.052	& 0.048 & \\	
			& \multirow{3}{*}{$1+\delta_1 t$} & 0 & 0.049 &	0.050 &	& 0.045 &	0.049 &	 \\
			& & 5 & 0.054 & & & 0.050 & & \\
			& & 10 & 0.053 & & & 0.051 & &
		\end{tabular}}
		\end{center}
\end{table}
Figure 1 presents power for the \textit{aRLRT} and \textit{aScore} methods for binary responses with densely observed functional predictors; Figure S3 in the Supplemental Materials compares all methods. While there is no uniformly best method, the \textit{aRLRT} method has similar or higher power than all other valid methods for all settings. The \textit{aScore} test has similar power to \textit{aRLRT} when $\beta(t)$ is scalar and linear, but can have significantly lower power for the trigonometric $\beta(t)$ when the null hypothesis is a poor approximate for the true coefficient. Power decreases as subject curves are more sparsely observed, particularly when the functional predictor is noisy and less accurately estimated (Supplemental Section 1.1). In some scenarios, the tests did not converge to 100\% power due to instability in the model estimates, as previously noted for type I error rates.
	
The \textit{aRLRT} method has higher power than the \textit{aPFR} and \textit{aFGAM} methods for all settings (Supplemental Table 2). In particular, \textit{aRLRT} has 5-10\% higher power for testing nullity when the competing methods require testing fixed and random effects, as a result of using the RLRT rather than LRT for testing. This demonstrates the benefit of an adaptive mixed model representation compared to existing static approaches. For binary responses, the \textit{FPCR} method for $n=100$ and the \textit{aFtest} for all settings are not valid because they do not maintain type I error rate. The \textit{FPCR} method is valid for $n=500$, and has comparable power for testing nullity but can have much lower power for testing functionality. Performance patterns for all methods are similar for testing Normal, Binomial, and Poisson responses, and power is generally higher for all methods when applied to Normal and Binomial data (Supplemental Sections 1.3-1.5). Thus, the \textit{aRLRT} method has best overall performance by (a) maintaining type I error rates close to the nominal level while (b) having consistently high power for all settings.
	
\begin{figure}
	\begin{center}
	\includegraphics[scale=0.14]{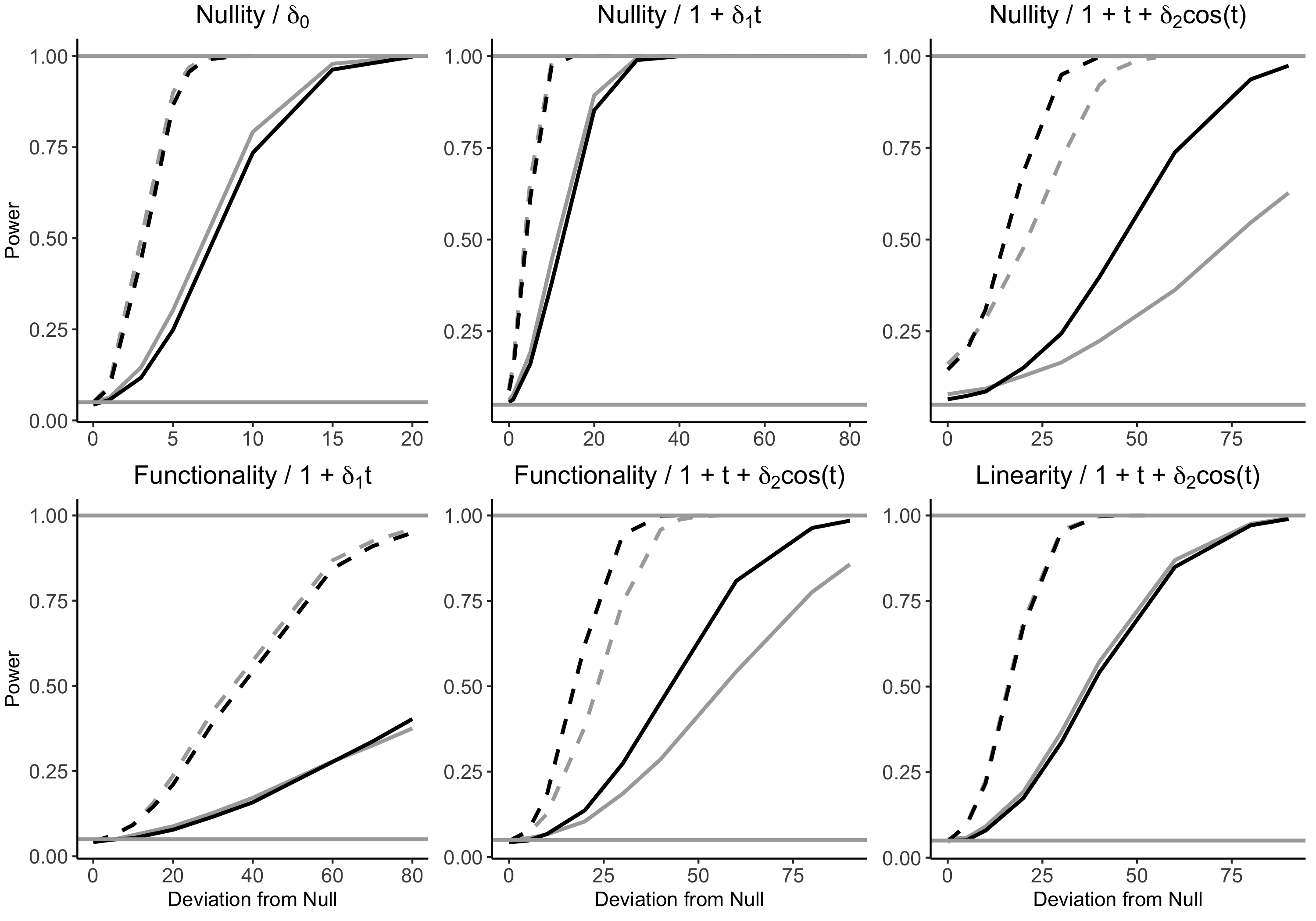}
	\caption{Power for dense binary data at the $\alpha=0.05$ level based on 5000 simulated datasets, by $\beta(t)$ form. Legend: \textit{aRLRT} (black), \textit{aScore} (gray), $n=100$ subjects (solid), $n=500$ subjects (dashed).}
	\end{center}
\end{figure}
	
\section{Applications}\label{sec:applications}
\subsection{Phoneme Classification}
We first consider an application to digital speech classification described by Hastie et al. (1995) in the context of penalized discriminant analysis, available as the \texttt{Phoneme} dataset in the \texttt{fds} package \citep{fds}. \cite{FerratyVieu2003} and \cite{MousaviSorenson2017} found that classification using functional approaches outperformed non-functional methods for curve discrimination. To formally test this observation, we apply the \textit{aRLRT} and \textit{aScore} methods to test the form of the smooth coefficient. We focus on the 400 phoneme curves for ``aa" (as in the vowel for ``dark") and 400 curves for ``ao" (as in the first vowel of ``water"). Each curve gives the log-periodogram as a function of frequency measured at a 16-kHz sampling rate, considering only the first 150 frequencies.
	
Figure 2 shows the data and estimated smooth coefficient from a scalar-on-function linear model with binary responses, where ``aa" is assigned value 0 and ``ao" is value 1. Visually, the smooth coefficient looks functional and non-linear. The \textit{aRLRT} method yields highly significant RLRT statistics for the tests for linearity, functionality, and nullity of 112.9, 145.2, and 175.7, respectively, corresponding to $p$-values $<0.001$. Similarly, the \textit{aScore} method yields highly significant statistics of 10406.6, 8755.5, and 1479.6, respectively, corresponding to $p$-values $<0.001$. Both methods indicate that the smooth coefficient has a non-linear functional form.
	
\begin{figure} 
	\begin{center}
	\includegraphics[scale=0.14]{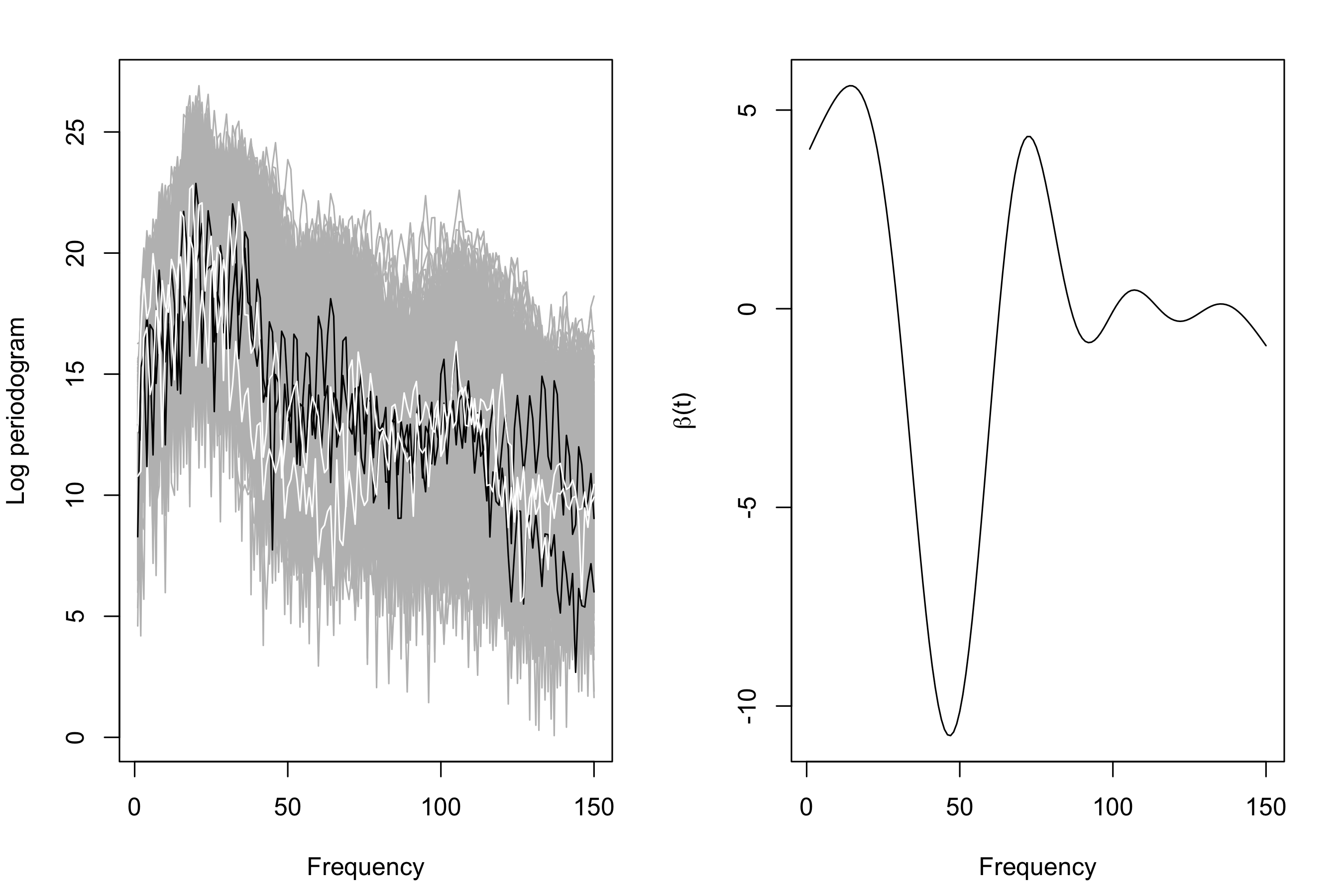} 
	\caption{(Left) Phoneme curves with two example ``aa" (black) and ``ao" (white) highlighted curves. (Right) Estimated coefficient function.}
	\end{center}
\end{figure}
	
\subsection{Identifying Multiple Sclerosis Patients using Diffusion Tensor Imaging}
We now consider the motivating example of identifying MS patients using DTI of intracranial white matter microstructure. We focus on the complete baseline fractional anisotropy tract profiles for the (a) corpus callosum (CCA), observed at 93 points for 42 healthy and 99 MS patients, and (b) right corticospinal tract (RCST), observed at 57 points for 26 healthy and 66 MS patients, shown in Figure 3. The CCA connects the right and left hemispheres of the brain and is associated with cognitive function, while the RCST connects to the spinal cord and is associated with motor function. We apply the \textit{aRLRT} and \textit{aScore} methods to determine the form of the smooth coefficient in scalar-on-function linear models.
	
\begin{figure} 
	\begin{center}
	\includegraphics[scale=0.14]{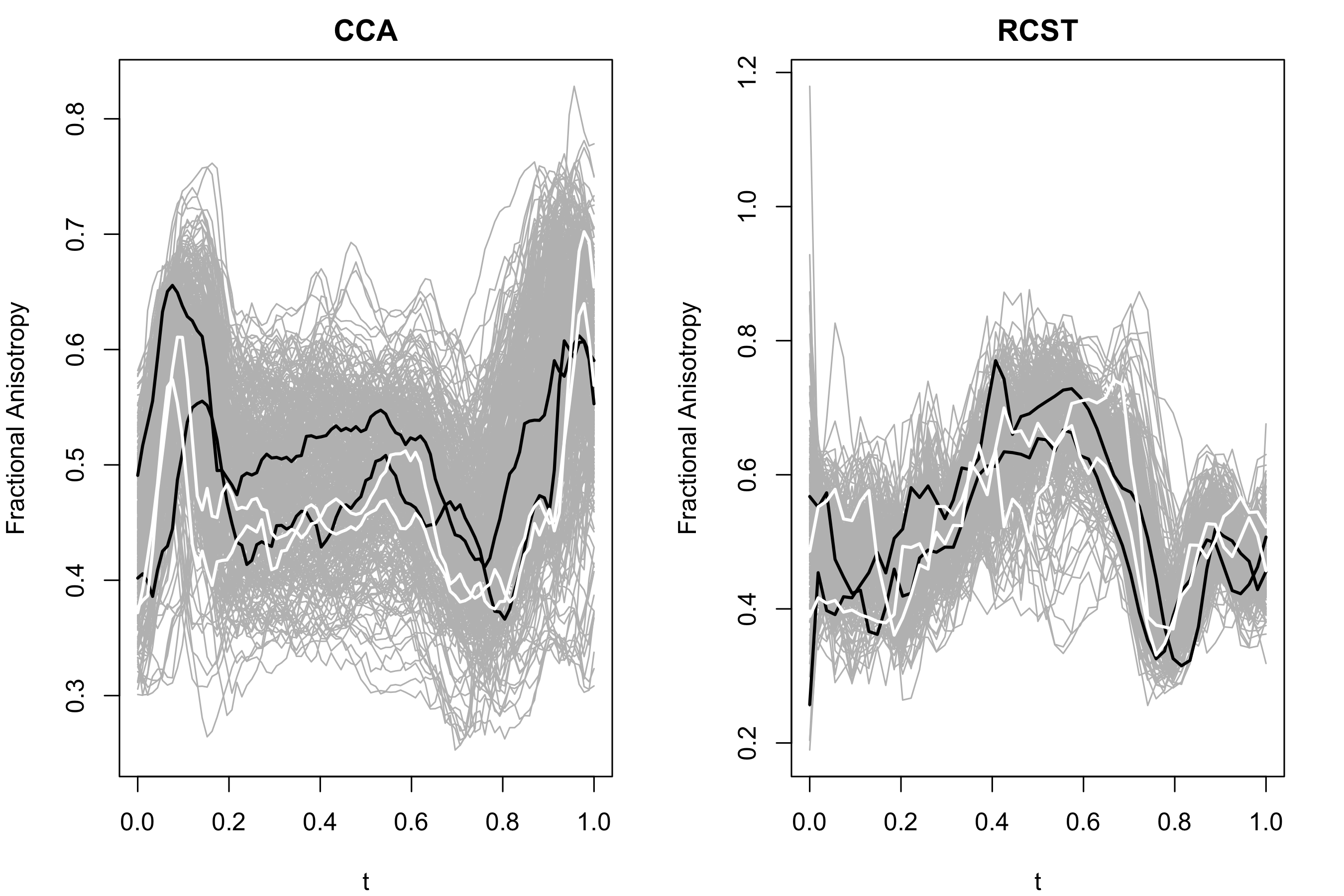}
	\caption{Baseline diffusion tensor imaging tracts for the corpus callosum (CCA) and right corticospinal tract (RCST) for healthy (black) and multiple sclerosis (white) patients.}
	\end{center}
\end{figure}
	
Table 2 reports results for the \textit{aRLRT} and \textit{aScore} methods. While both methods indicate that CCA tracts have a significant non-zero relationship while RCST tracts are unrelated to MS, the simulation study showed that both methods could have low power for binary data with few subjects and weak signal, as in these datasets. In the next section, we consider a power analysis of the CCA data to determine if the methods are underpowered for this application.
	
\begin{table}
	\begin{center}
	\caption{Testing results for the diffusion tensor imaging (DTI) dataset using the \textit{aRLRT} and \textit{aScore} methods. }
	\scalebox{0.9}{\begin{tabular}{ c | c | c c c | c c c }
		& & \multicolumn{3}{c|}{\textit{aRLRT}} & \multicolumn{3}{c}{\textit{aScore}} \\ \hline
		& & Linearity & Functionality & Nullity & Linearity & Functionality & Nullity \\\hline
		\multirow{2}{*}{CCA} & statistic & 0.829 & 0.658 & 24.381 &
		0.205 & 0.022 & 0.058 \\
		&$p$-value & 0.094 & 0.126 & $<$0.001 & 0.120  & 0.176 & $<$0.001 \\\hline
		\multirow{2}{*}{RCST} & statistic & 0.647 & 0.786 & 1.030 & 
		0.298 & 0.024 & 0.002 \\
		& $p$-value & 0.111 & 0.119 & 0.108 & 0.094 & 0.122 & 0.159
		\end{tabular}}
	\end{center}
\end{table}
		
\subsubsection{Power Analysis}
Simulated datasets are generated based on the baseline CCA scans of both healthy and MS patients using the estimated $\mu(t)$, $\lambda_k$, $\sigsq_e$, and $K_x=31$, as estimated by AIC, as done in the simulation study in Section 5. We set the smooth coefficient as $\beta(t)=\delta \hat{\beta}(t)$, where $\delta$ controls the magnitude of $\beta(t)$ and $\hat{\beta}(t)$ is the estimated smooth coefficient from the test for linearity (Figure 4, top left panel). Note that $\delta=0$ is the null hypothesis for all three hypotheses and $\delta=1$ is the estimated smooth coefficient. The estimated smooth coefficient looks approximately linear or quadratic. Because both the \textit{aRLRT} and \textit{aScore} methods may be underpowered for the given sample size, we also consider datasets with double the number of subjects.
		
\begin{figure}
	\includegraphics[scale=0.16]{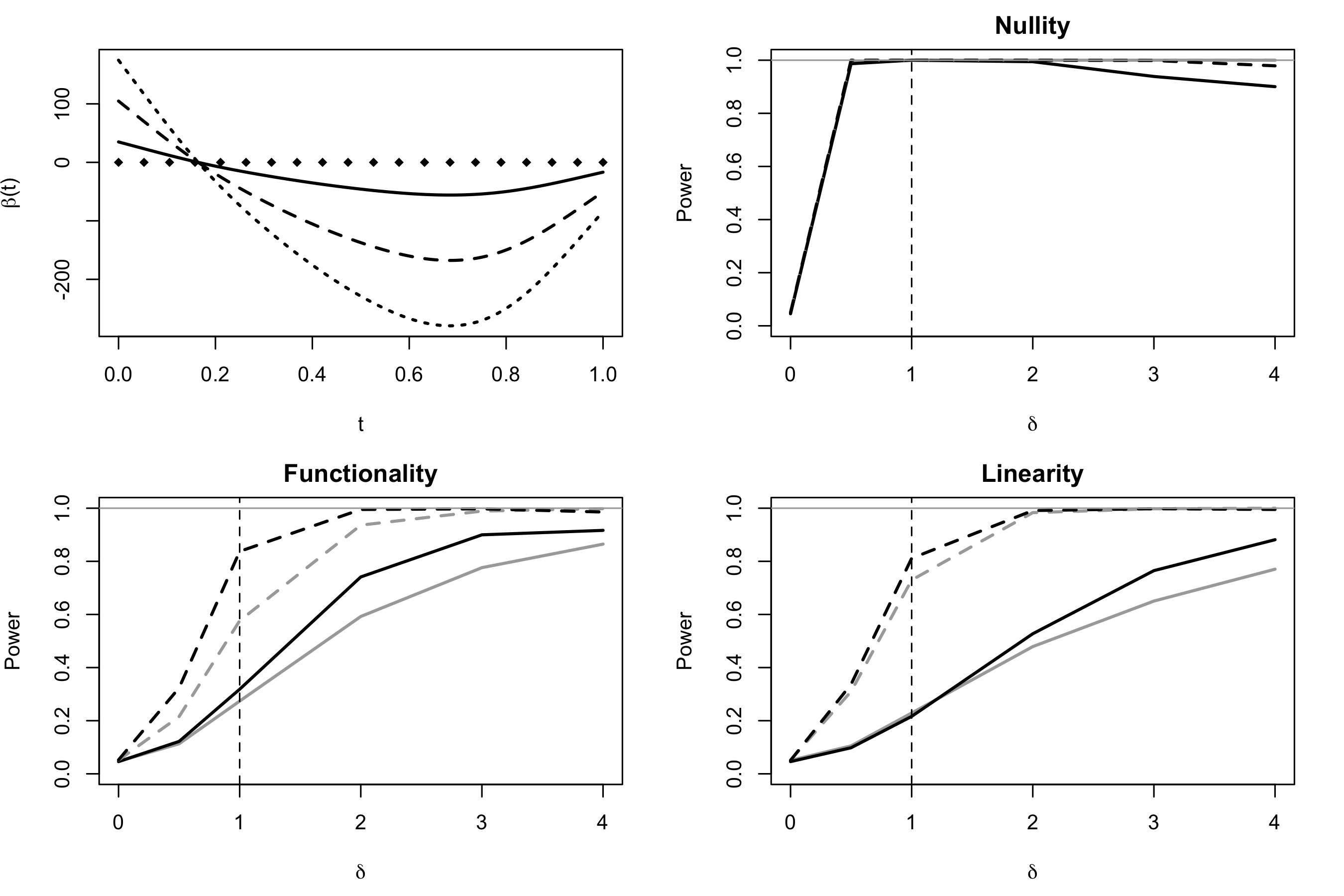}
	\caption{(Top left) Example $\beta(t)=\delta \hat{\beta}(t)$, where $\delta$ controls magnitude and $\hat{\beta}(t)$ is the estimated smooth coefficient. Legend: $\delta=0$ (dotted), $\delta=1$ (solid), $\delta=3$ (long dash), $\delta=5$ (short dash). (Remaining panels) Power for simulated data at the $\alpha=0.05$ level based on 5000 simulated datasets for tests for nullity, functionality, and linearity. Legend: \textit{aRLRT} (black), \textit{aScore} (gray), standard dataset (solid), data with $\times 2$ subjects (dashed).}		
\end{figure}
		
From Figure 4 it is clear that while only the test for nullity has high power for the true sample size, doubling the number of subjects is sufficient to achieve $>80$\% power for the \textit{aRLRT}. This analysis also shows that for small sample sizes, power can decrease even when deviation from the null increases due to instability in the model estimates. Although both methods are likely underpowered for the true dataset, this power analysis suggests that the coefficient may be functional and non-linear. This indicates a non-linear relationship between fractional anisotropy and MS disease status.
		
\section{Conclusion}
We propose an approximate restricted likelihood ratio test framework for the smooth coefficient in scalar-on-function linear models with generalized responses. This test can be used compare functional and non-functional linear models with responses from any exponential family distribution. Our method performs well compared to several competitors for dense and sparse data from Bernoulli, Poisson, Binomial, and Normal distributions. Caution should be taken when estimating and testing models for Bernoulli responses when the estimated probabilities are extreme. We apply our test to classifying phoneme curves and identifying multiple sclerosis patients using diffusion tensor imaging. 

\bigskip
\begin{center}
{\large\bf SUPPLEMENTARY MATERIAL}
\end{center}

\begin{description}
\item[Additional Results:] Additional simulation results for sparse data, extended competitor methods, and other exponential family distributions available on request from corresponding author.

\item[R code:] R code for all six methods considered in this paper available on request from corresponding author.
\end{description}

\bibliographystyle{Chicago}

\bibliography{GFLMbib}
\end{document}